Fabrication of binary FeSe superconducting wires by diffusion process

Toshinori Ozaki[1,3], Keita Deguchi[1,2,3], Yoshikazu Mizuguchi[1,2,3], Yasuna Kawasaki[1,2,3], Takayoshi Tanaka[1], Takahide Yamaguchi[1,3], Hiroaki Kumakura[1,2,3], and Yoshihiko Takano[1,2,3]

[1] National Institute for Materials Science, 1-2-1 Sengen, Tsukuba, Ibaraki 305-0047, Japan

[2] University of Tsukuba, 1-1-1Tennnodai, Tsukuba, Ibaraki 305-0047, Japan

[3] JST, Transformative Research-project on Iron Pnictides, 1-2-1 Sengen, Tsukuba, Ibaraki 305-0047, Japan

(Abstract)

We report successful fabrication of multi- and mono-core FeSe wires with high transport critical current density $J_c$ using a simple *in-situ* Fe-diffusion process based on the powder-in-tube (Fe-diffusion PIT) method. The seven-core wire showed transport $J_c$ of as high as 1027 A/cm$^2$ at 4.2 K. The superconducting transition temperature $T_c^{zero}$ was observed at 10.5 K in the wire-samples, which is about 2 K higher than that of bulk FeSe. The Fe-diffusion PIT method is suitable for fabricating multi-core wires of the binary FeSe superconductors with superior properties.

## I. INTRODUCTION

Since the discovery of the superconductivity in LaFeAsO$_{1-x}$F$_x$[1], several types of iron-based superconductors with layered structure have been discovered[2-5]. Among these iron-based superconductors, tetragonal FeSe with transition temperature of $T_c^{zero}$ ~8 K and $T_c^{onset}$ ~12 K has the simplest structure (PbO-type) and binary composition, consist of a stack of FeSe layers along the $c$-axis[4,6]. The starting materials for FeSe are less toxic compared to the As-based compounds, making it potential candidate for practical applications. It is also know that the $T_c$ of Fe chalcogenide is quite sensitive to compressive strain. The $T_c$ in FeSe is increased up to 37 K under high pressure[7,8]. The wires of Fe chalcogenide, which show increased $T_c$ by applying compressive strain, must be the greater advantage of wire applications.

Several attempts of wire fabrication for practical applications have been carried out in Fe-based compound[9-13]. We succeeded in observing transport $J_c$ for FeSe$_{1-x}$Te$_x$ superconducting wire[9,13]. In contrast, there have been few reports about the fabrication of the FeSe wire showing transport $J_c$[12]. Here we report the observation of the transport $J_c$ in multi- and mono-core wires of FeSe fabricated using a simple *in-situ* Fe diffusion powder-in-tube (Fe-diffusion PIT) method. Unlike the bulk synthesis process[4,6], this method involves only one thermal treatment. The interesting aspect of this process is that Fe sheath plays the role of not only the sheath but also the raw materials for synthesizing the superconducting phase. For synthesizing FeSe phase, the effective diffusion-distance of Fe needs to be shortened so that Fe reacts with Se before it evaporates. The multi-core wires prepared by this method could be more efficient to obtain higher $J_c$, because the diffusion-distance of Fe in the individual wires are further shorter, leading to better possibility for the reaction without the escape of Se. The observed transport $J_c$ for the seven-core wire was 1027 A/cm$^2$ at 4.2 K. Furthermore, our process produced a $T_c^{zero}$ of 10.5 K which about 2 K higher than that of bulk samples.

## II. EXPERIMENTAL

The Se powder was packed into pure Fe tubes with a length of 48 mm. The inner and outer diameters of the Fe tubes were 3.5 and 6.2 mm, respectively. The tubes were rolled into a rectangular rod of ~2.5 mm in size using groove rolling. After rolling, they were drawn into a wire of 1.1 mm in diameter using wiredrawing die. These wires were cut into pieces of ~5 cm in length. Some of these pieces were used as samples of mono-core wires. The seven-core wires were produced by packing the unsintered seven pieces of the mono-core wires into another Fe tube. The seven-core composites were drawn down to final diameter of 2.0 mm. The seven-cores wires were also cut in ~5 cm long pieces. These mono- and seven-core wires were sealed inside a quartz tube with argon gas. These sealed wires were taken into a furnace heated at 800 °C, held at this temperature for 2 hours, and then taken out from furnace to quench them.

The microstructure of these wires was investigated with scanning electron microscope (SEM) and x-ray diffraction (XRD). The surface mapping analysis of the wire was carried out using energy dispersive x-ray spectrometry (EDX). Transport critical currents ($I_c$) were measured for 4-cm-long wires by a standard four-probe resistive method in liquid helium (4.2 K) and applied magnetic fields. The magnetic field was applied perpendicularly to the wire axis. The criterion of $I_c$ definition was 1 $\mu$V/cm. The $J_c$ was obtained by dividing $I_c$ by the cross sectional area of the FeSe core excluding the hole, which was measured by optical microscope.

## III. RESULTS AND DISCUSSION

Figure 1(a) and 1(b) show respectively the polished transverse cross section of FeSe mono- and seven-core wires after heat treatment at 800°C for 2 hours. The FeSe layer was observed on the inside wall of the Fe sheath, and a hole was formed at the center of each core where the Se powder was filled before the heat treatment.

Figure 2 shows the SEM image and elemental mapping images for polished longitudinal cross section of the mono-core FeSe wire. As can be seen in the SEM image, FeSe layer is dense and monolithic with no reaction layer between the superconducting core and the sheath, suggesting the good connection between them. The elemental mapping analysis showed that the Fe distribution is homogeneous in the superconducting phase, which indicates that the Fe sheath reasonably supplied Fe for synthesizing superconducting phase of FeSe. The dispersion of Se is also homogeneous. These results indicate that the FeSe phase inside the Fe sheath was expectedly synthesized by the Fe-diffusion PIT method.

Figure 3 shows the XRD pattern of reacted layer obtained from the mono-core wire.

It is found that the main peaks were well indexed on the basis of the tetragonal PbO-type structure with the space group of P4/*nmm*, and the minor peaks were identified as iron-oxide and hexagonal phase. Lattice constants were calculated to be $a$ = 3.7689(9) and $c$ = 5.5023(32) Å. Obtained lattice parameter $c$ of the wire is slightly smaller than the value of 5.520(1) Å for bulk FeSe[14], indicating a lattice compressive strain along $c$-axis. This compressive strain may be resulted from the quench of wire in the heat-treatment process.

Temperature dependence of resistivity for the mono-core FeSe wires under different applied magnetic fields is shown in Fig. 4. Interestingly, the resistivity at 0 T began to decrease at 12.3 K and drops to zero at 10.5 K. The $T_c^{zero}$ is ~2 K higher than that of FeSe reported in bulk samples[4,6,15]. The similar effect was reported in the FeSe$_{0.5}$Te$_{0.5}$ films[16,17] and Fe$_{1.03}$Se synthesized by flux method[18]. It is also reported in our previous report[19] that the $T_c$ of the iron-based superconductor was strongly correlated with the anion height from Fe layer. Given these facts, it would be understood that the enhancement of $T_c^{zero}$ in FeSe wire should relate to the shrinkage of $c$-axis value, arising from a compressive strain. The $\rho(T)$ curves are shifted to lower temperatures with increasing magnetic fields without noticeable broadening compared to the zero-field case. The transition width $\Delta T$ defined by the 90 and 10% point on $\rho(T)$ is less than 2 K. This behavior is similar to that of the low-temperature superconductor with small anisotropy[20,21]. The inset of Fig. 4 shows the temperature dependence of upper critical field ($\mu_0H_{c2}$) and the irreversibility field ($\mu_0H_{irr}$) determined by using criteria of 90% and 10% drop of normal state resistivity. The $\mu_0H_{irr}$ line is very close to the $\mu_0H_{c2}$ line. Linear extrapolation of the $\mu_0H_{c2}(T)$ and $\mu_0H_{irr}(T)$ data suggests $\mu_0H_{c2}(0) \sim 32$ T and $\mu_0H_{irr}(0) \sim 23$ T.

The transport $J_c$ as a function of magnetic fields for FeSe wires at 4.2 K is presented in Fig. 5. We succeeded in observing the transport $J_c$ for both multi- and mono-core FeSe wires. The mono-core wire showed a transport $J_c$ of 350 A/cm$^2$ at 4.2K. Furthermore, the transport $J_c$ for seven-core wire reached as high as 1027 A/cm$^2$ at 4.2 K. The transport $J_c$ for seven-core wire showed an enhancement by a factor of ~10 compared to that in the previous report for FeSe wire[12]. This high $J_c$ value would results from an enhancement of grain connectivity, due to the higher sintering temperature used in the present synthesis process. The $J_c$ of FeSe wires gradually decreased with increasing magnetic field up to 12 T, indicating that the FeSe wires have clear advantages for the wire applications under high magnetic fields. Our result demonstrates that the Fe-diffusion PIT method is greatly effective for fabricating the multi-core FeSe wires. We expect that much higher $J_c$ could be realized by further improving grain-boundary conductivity and reducing the amounts of inclusions such as iron-oxide and hexagonal phase.

## IV. CONCLUSION

We fabricated seven- and mono-core wires of FeSe using the Fe-diffusion PIT method with Fe sheath. The Fe-diffusion PIT method is the simplest process of all the wire-fabrication processes. We have succeeded in synthesizing high quality FeSe superconducting phase inside the Fe sheath. The seven-core superconducting wires showed a transport $J_c$ as high as 1027 A/cm$^2$. The relatively high $T_c^{zero}$ = 10.5 K was obtained for the FeSe wire, which might be attributed to the shrinkage of the $c$-axis. These results show that the Fe-diffusion PIT method is suitable for fabricating iron-based superconducting multi-core wires with higher $J_c$ and $T_c$.

"Acknowledgments"

This work was supported in part by the Japan Society for the Promotion of Science (JSPS) through Grants-in-Aid for JSPS Fellows and 'Funding program for World-Leading Innovative R&D on Science Technology (FIRST) Program'.

(Captions)

Fig. 1 Cross section view of (a) mono- and (b) seven-core wires of FeSe after heat treatment, which shows the Fe sheath, FeSe layer and the void space in the center.

Fig. 2 SEM image and elemental mapping on the longitudinal cross section for the mono-core FeSe wire fabricated by the *in-situ* Fe-diffusion PIT method.

Fig. 3 XRD pattern for FeSe superconducting wire fabricated by the *in-situ* Fe-diffusion PIT method.

Fig. 4 Temperature dependence of resistivity for mono-core wires fabricated by the *in-situ* Fe-diffusion PIT method under magnetic fields up to 7 T. The inset shows temperature dependence of $\mu_0 H_{c2}$ and $\mu_0 H_{irr}$ determined from 90% and 10% points on the resistive transition curve.

Fig. 5 Magnetic field dependence of transport $J_c$ at liquid helium temperature (4.2 K)

for mono- and seven-core FeSe wires fabricated by the *in-situ* Fe-diffusion PIT method.

The magnetic field was applied perpendicular to the wire axis.

Figure 1

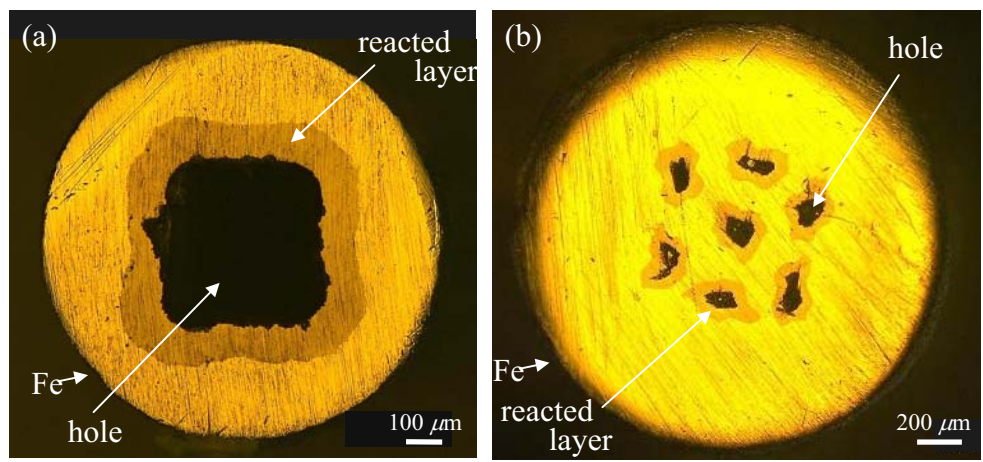

Figure 2

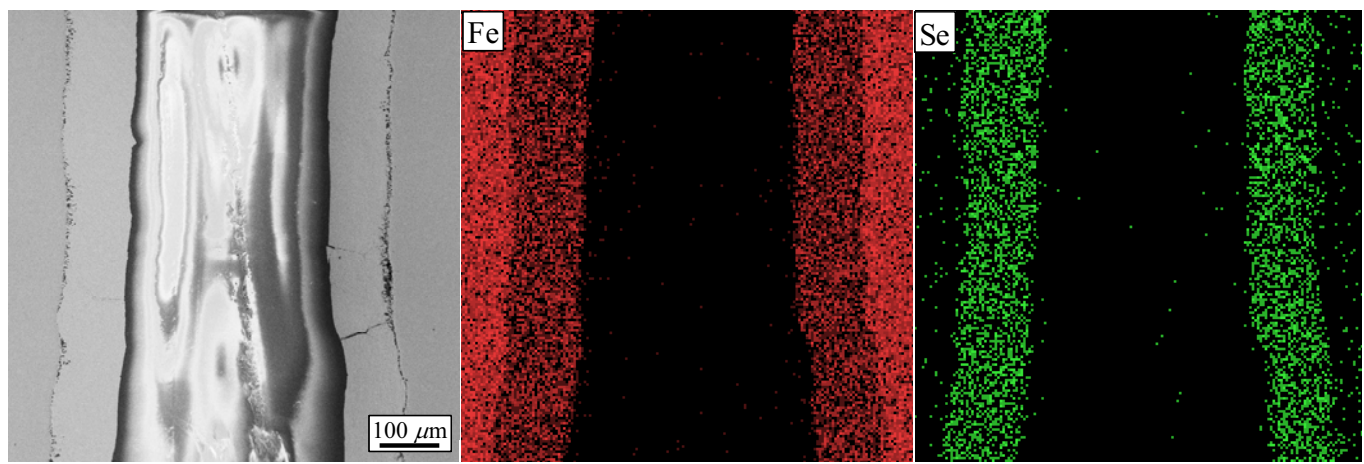

Figure 3

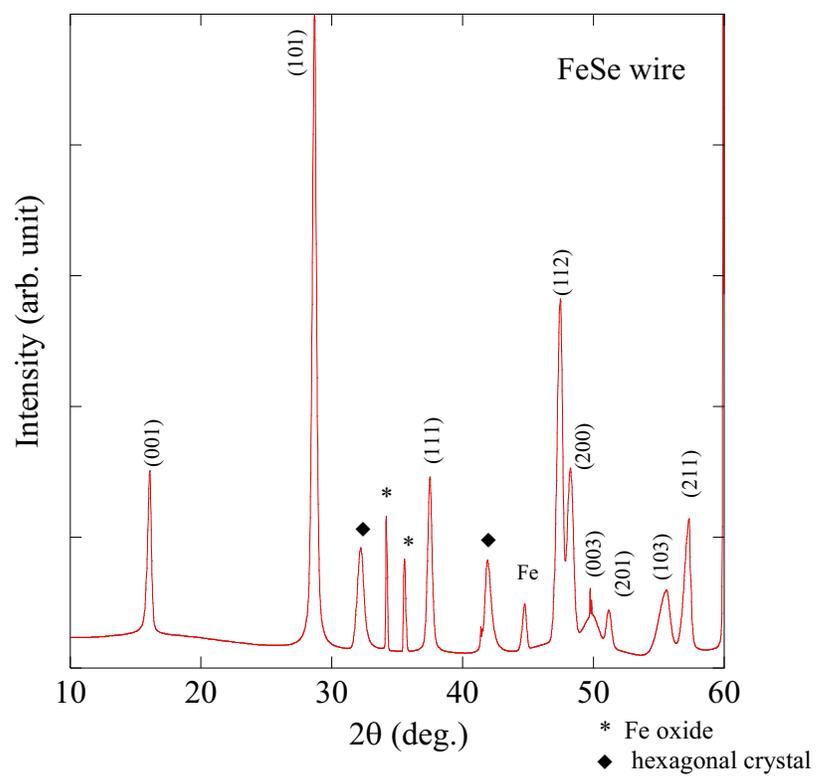

Figure 4

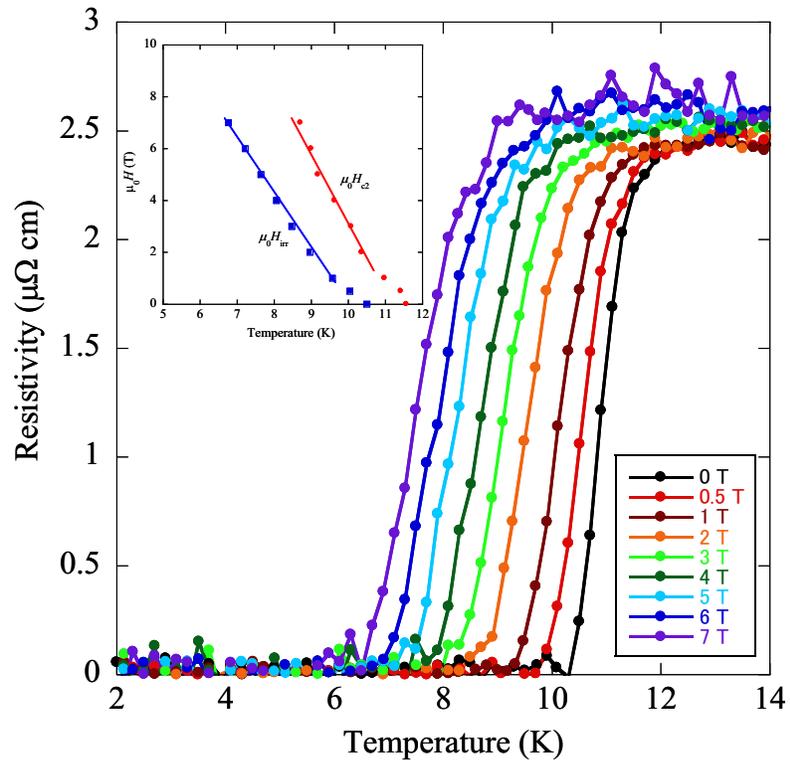

Figure 5

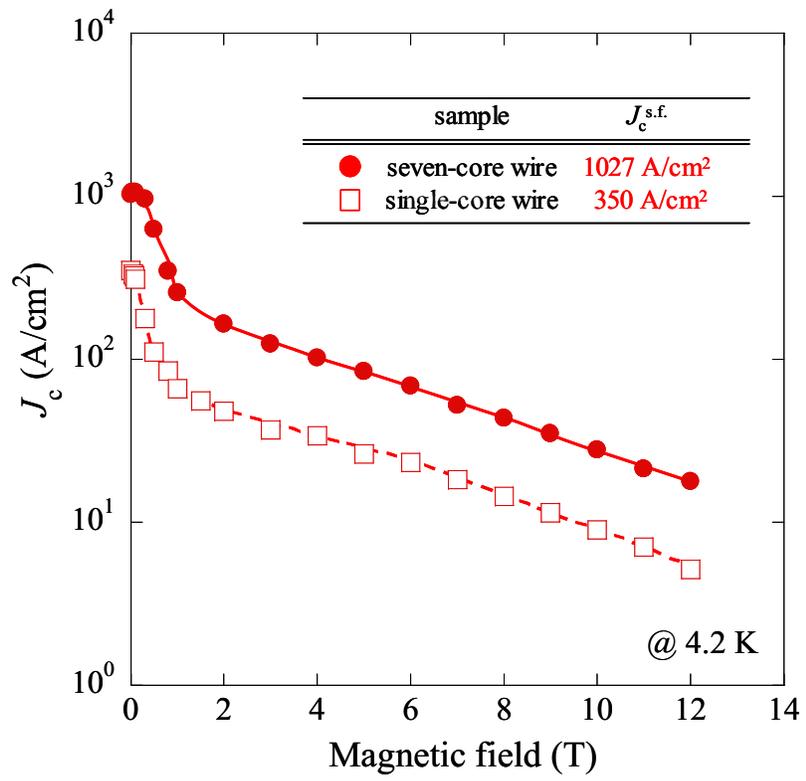